Continuously control of polarization via electrically driven long-distance superlubric sliding

Peiyao Shi and Menghao Wu*

School of Physics, Huazhong University of Science and Technology, Wuhan, Hubei 430074, China

**Accessible overview**

Superlubric sliding of incommensurate interfaces with ultra-low barriers has garnered significant interest, while its potential is greatly limited since it was mechanically driven by tips in previous studies. This work proposes a design of heterojunctions with continuously tunable vertical polarizations via electrically driven superlubric sliding, rendering series of multi-states for artificial synaptic devices. The polarization switching barriers are unprecedentedly reduced to the magnitude of μeV, with ultra-long ion displacements distinct from current paradigm of ferroelectricity, providing significant scientific and technological opportunities.

**Highlights**

· The polarization switching barriers are unprecedentedly reduced to the magnitude of μeV.

· Superlubric sliding electrically driven by low vertical voltage is hitherto reported, much advanced compared with mechanical driven approach.

· The ultra-long ion displacements are distinct from current paradigm of ferroelectricity, rendering continuous control of multi-states for artificial synaptic devices.

**Summary**

Sliding ferroelectricity widely exists in various van der Waals bilayers/multilayers, which is induced by asymmetric stacking of commensurate interface. The greatly reduced switching barriers via interlayer sliding lead to high speed with low energy cost, while they are still much higher compared with superlubric sliding of incommensurate interfaces. The polarizations of such incommensurate interfaces are not switchable, which is the major obstacle of combing superlubricity and sliding ferroelectricity for ultralow barriers. Here we propose a design of such combination based on previous synthesis of lateral heterojunctions of 2D materials, which can be extensively applicable to various systems including PN junctions. In such long-distance superlubric ferroelectricity, the vertical polarization can be continuously controlled by superlubric sliding of incommensurate interfaces between lateral heterojunction bilayers, where the series of multiple stable states are long-sought for artificial synaptic devices. The unconventionality of our findings does not only include unprecedented barriers down to the magnitude of $\mu$eV, but also unprecedented long ion displacements distinct from the small deviations in classical paradigm of ferroelectricity. Our predicted superlubric sliding electrically driven by low vertical voltage is also hitherto reported, much more efficient compared with previously reported sliding mechanically driven by tips, resolving a major issue for practical applications.

**Graphical abstract**

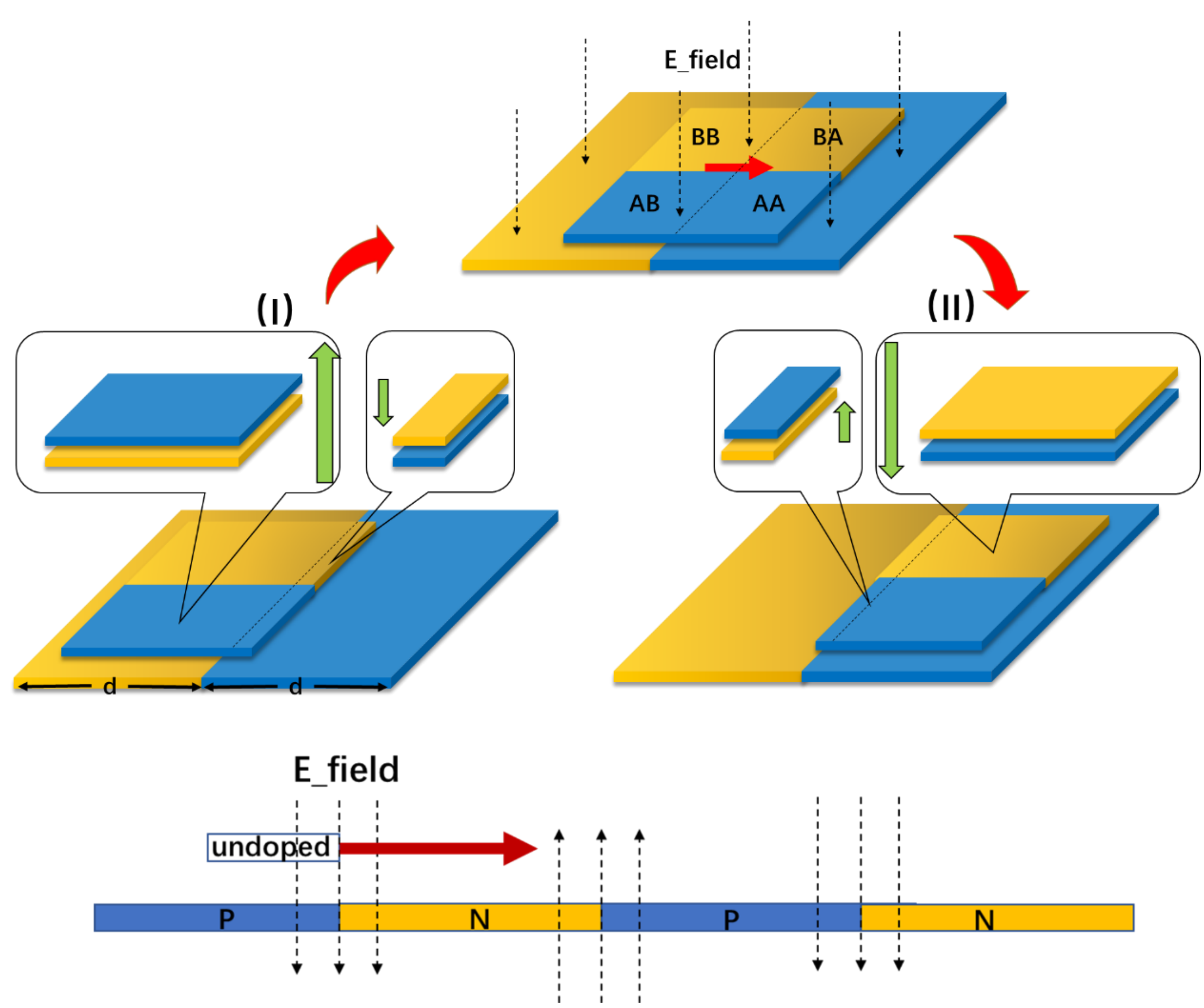

## Introduction

Ferroelectric switching typically involves small ion displacements from paraelectric phase. As proposed by Abrahams, no atom in the unitcell should be displaced more than about 1 Å along the polar direction.[1] For example, the Ti ions of $BaTiO_3$ in the ground state only deviate from the centers of $TiO_3$ octahedrons by less than 15 pm. Herein longer ion displacements generally imply higher switching barriers, which may be favorable for stability but undesirable for achieving high speed by low switching voltage. Currently, the lowest switching barriers known in room-temperature ferroelectricity exist in two-dimensional (2D) materials. Even for most 2D non-polar monolayers with high symmetry crystal lattices, vertical polarizations can be formed in their bilayers/multilayers upon asymmetric stacking. They are switchable via interlayer sliding with low barriers[2] (around the scale of meV, still higher compared with the magnetic anisotropy of most magnets), while the thermal stability can still be ensured due to the intralayer rigidity.[3] Within several years, such so-called sliding ferroelectricity has been experimentally confirmed in various van der Waals systems,[4-14] where high-speed and fatigue-resistant devices have also been realized,[15-18] rivaling state-of-the-art ferroelectrics.

The interlayer interfaces of sliding ferroelectrics are commensurate. The macroscopic ferroelectricity will disappear when the stacking becomes incommensurate upon a large twist angle, while the sliding barriers and frictions almost vanish due to the offset of atomic lateral force. Such structural superlubricity with negligible frictions is of great significance for maintaining service life and conserving energy in mechanical systems, which has been experimentally demonstrated in various van der Waals incommensurate contacts like graphene (tip)/graphite[19], graphene-$MoS_2$,[20] graphene-hBN[21], $MoS_2$/graphite, and $MoS_2$/h-BN heterojunctions[22]. Superlubric sliding in previous studies was mechanically driven by atomic force microscope (AFM) tips, which can scarcely be driven by electric field, so the potential applications are greatly limited. Even superlubric sliding of less than 2 Å under an electric field has been predicted in sandwiched trilayer ferroelectric systems with commensurate across-layer configurations,[23] electrical driven sliding by long distance remains to be a challenge. Although electrostatic in-plane superlubric actuator has been realized utilizing charge injection in a recent report[24], a high in-plane voltage around 200V is required. In this paper, we propose a design that enables long superlubric sliding driven by a low vertical voltage, which can control the polarization continuously. The switching barriers are unprecedentedly reduced to the magnitude of μeV, orders of magnitude lower compared with previously studied sliding ferroelectrics and even lower than the magnetic anisotropy of most magnets, with ultra-long ion displacements distinct from classical ferroelectricity.

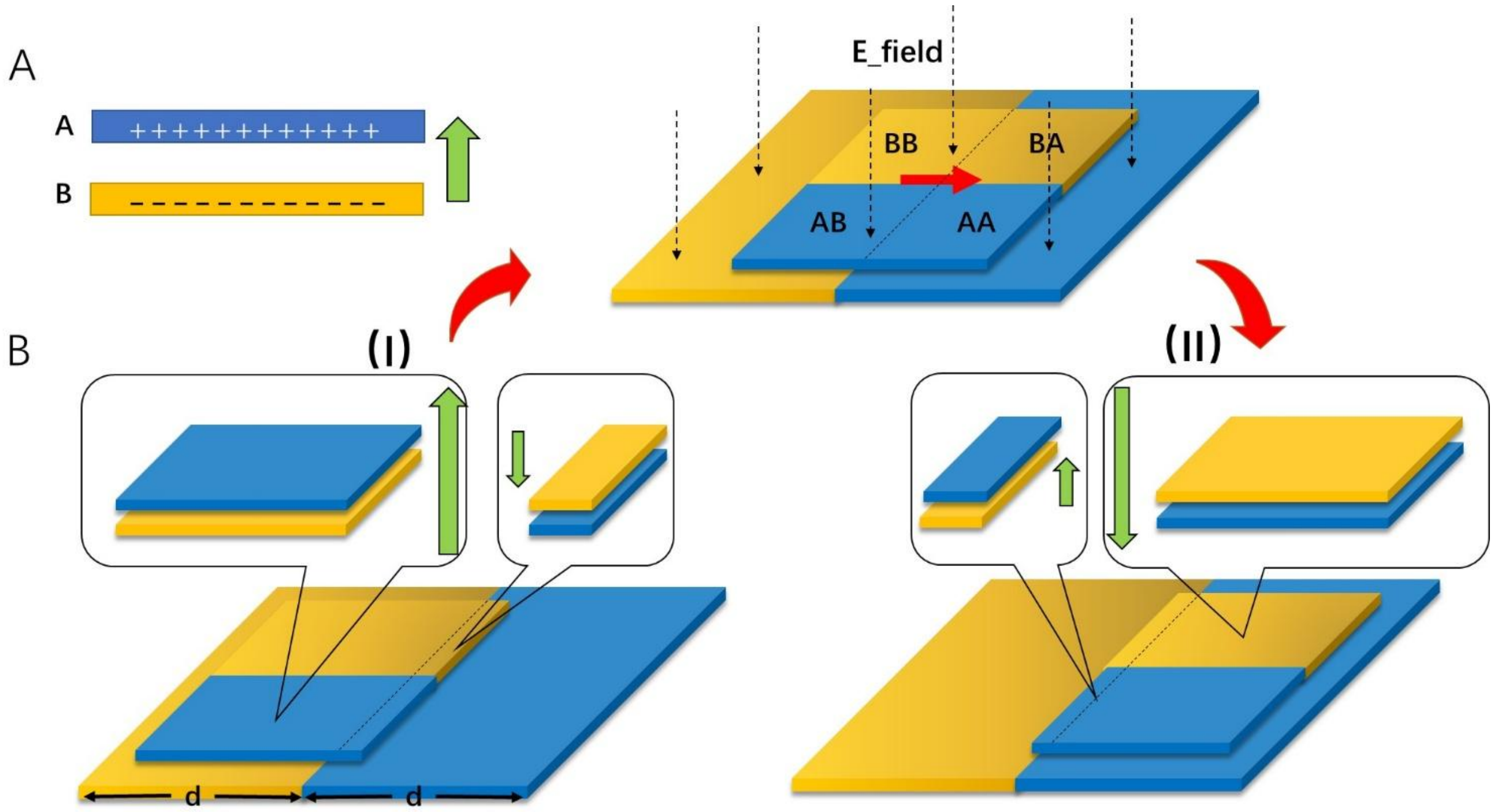

Figure 1 (A) Charge transfer and vertical polarization in AB hetero-bilayer. (B) The design of superlubric device based on bilayer AB lateral junctions, where polarization (marked by green arrows) switching is realized via superlubric sliding under a vertical electric field changing the ratio of AB/BA regions.

## Results

It is not a difficult task to form vertical polarizations in heterobilayers, while the major challenge is polarization switching. As shown in Fig. 1A, when two monolayers of different materials are stacked together, a vertical polarization will be induced by interlayer charge transfer, while such charge transfer is not reversible. Aside from vertical heterobilayers, various laterally stitched heterostructures including graphene(Gr)/BN,[25-27] Gr/$MoS_2$,[28-30] Gr/$WSe_2$,[31], $MoS_2$/$WSe_2$,[32] $MoSe_2$/$WS_2$, $MoS_2$/$WS_2$, $WS_2$/$WSe_2$, etc.,[33] have been realized via scalable and patternable growth. Here we propose a design based on such heterojunctions.

Suppose a monolayer rectangular heterojunction flake of size d×l composed of two different materials A and B, is placed on a similar monolayer AB heterojunction flake of larger size 2d ×L, with the AB boundaries of two flakes perpendicularly crossing, as shown in Fig.1B. Now the overlapping regions can be divided into non-polar AA/BB domains and AB/BA regions with polarizations of opposite directions. For configuration (I) with right edge of the top flake aligned with the A/B boundary of the bottom flake, the stacking region is composed of polar AB stacking regions (with polarization $P_0$ upwards) and non-polar BB stacking regions. As the top flake slides towards right by length d, the polar AB region becomes non-polar AA region in configuration (II), while the non-polar BB stacking region is transformed to polar BA region (with polarization $P_0$ downwards), so the average switching polarization of the bilayer overlapping region $P=(S_{AB}-S_{BA})P_0/S_{total}=P_0/2$, where $S_{AB}$, $S_{BA}$, $S_{total}$ denote the area of AB, BA, and the total area of the top flake.

Upon a vertical electric field covering the whole system, it is more favorable in energy for the polarization to be aligned along the same direction, so such interlayer sliding can be electrically driven similar to typical sliding ferroelectricity. Meanwhile the interface can be incommensurate upon 90 degree twist angle between the bottom and upper layers, so the sliding barrier can be greatly reduced. Moreover, all intermediate states in the sliding pathway are stable, where the average polarization $P=(S_{AB}-S_{BA})P_0/S_{total}$ is tunable within the range $[-P_0/2, P_0/2]$. The charge density and conductance will vary with the vertical polarization upon sliding depending on the duration time and voltage, and such continuously change for nonvolatile multilevel resistance[34, 35] in mimicking synaptic functions are long-sought for constructing artificial neural systems.

The synthesis of monolayer AB junction is a prerequisite. As a paradigmatic case in our study, the lateral heteroepitaxy of graphene/BN monolayer with similar lattice has been realized and developed.[25-27, 36] The vertical polarization $P_0$ of twisted graphene/BN heterobilayer with incommensurate interface (approximately treated in a large unitcell in Fig. 2A) is estimated to be 0.38 pC/m. If the superlubric heterobilayer device is fabricated based on graphene/BN junction, the polarization of overlapping region $P=(S_{AB}-S_{BA})P_0/S_{total}$ can be tuned between -0.19 pC/m and 0.19 pC/m, as shown in Fig. 2B. Meanwhile the sliding barrier can be reduced to μeV scale as revealed by the calculated sliding pathway of the upper layer for one lattice constant along –x direction in Fig. 2C, which also accords with previous studies of superlubricity in graphene and BN systems. [21, 37-39] It is noteworthy that almost all the experimentally reported sliding ferroelectricity are robust at ambient conditions despite their low switching barriers, which are "collective" barriers where all dipoles must "simultaneously" switch to the opposite direction.[40] The in-plane rigidity makes the dipole disorder state highly unfavorable in energy (i.e., high "isolated" barrier), and such mechanism of high thermal stability can also be applicable to our design, where the robustness can be ensured as long as ~$k_B$T cannot overcome the total barrier as a whole. To our calculations, for a Gr/BN flake with size larger than 10 nm×10 nm, the total sliding barrier will be over $k_B$T at 300K. For its application in synaptic devices, the structural reconstruction and pinning dominant in moire systems are negligible in our incommensurate van der Waals system with large twist angle, and its reliability and stability can be ensured when the top slider flake is larger than this size.

In previous reports,[6, 41] sliding ferroelectricity in metallic thin-layer $WTe_2$ has been experimentally confirmed, which is attributed to the vertical confinement of electrons in the thin-layer $WTe_2$ even it is metallic in-plane,[40] so the vertical polarization will not be fully screened and can be switchable. Similarly, in graphene-based heterobilayer, the electron is vertically confined in graphene monolayer. For the G/BN heterobilayer upon a vertical electric field of 0.1 V/Å, the corresponding change in planar-averaged potential along the z direction is displayed in Fig. S1, which is estimated using standard method for DFT calculations of electrostatics.[42, 43] If the interlayer distance is coarsely taken as the effective thickness of BN and graphene monolayer, by calculating the chemical potential difference $\Delta V$, the average penetrated electric field $\Delta V/\Delta d$ can be obtained, which is respectively 0.047 and 0.041 V/Å for BN and Gr layer, both over 40% of the applied external electric field, close to the value of 37% obtained in graphene bilayer.[42] Actually, there have already been many experimental reports on graphene-based ferroelectrics recently, where the electrical-driven sliding seems to not have been impeded by charge screening.[44-46]

We may estimate the upper limit of the required field E = $\Delta_{barrier}/\Delta P$. where $\Delta_{barrier}$ is the barrier for one lattice constant, and ΔP is the corresponding change of polarization upon such displacement. It turns out E=0.005d V/$Å^2$ depends on d, so for d=2 nm, the required electric field and voltage will be respectively 0.1 V/Å and 0.5 V if the bilayer thickness is around 5 Å. However, the coercivity field of ferroelectrics is generally overestimated by orders of magnitude using such coarse method regardless of temperature and boundaries. For example, the theoretical coercivity electric field for $BiFeO_3$ (with switching barrier of 0.45 eV/f.u. and polarization around 90 μC/$cm^2$)[47] will be around 10 MV/cm, more than 2 orders of magnitude higher compared with the experimental value.[48] In this regards, it is likely that slider flake of much larger size can actually be driven by much lower voltage compared with our predicted values.

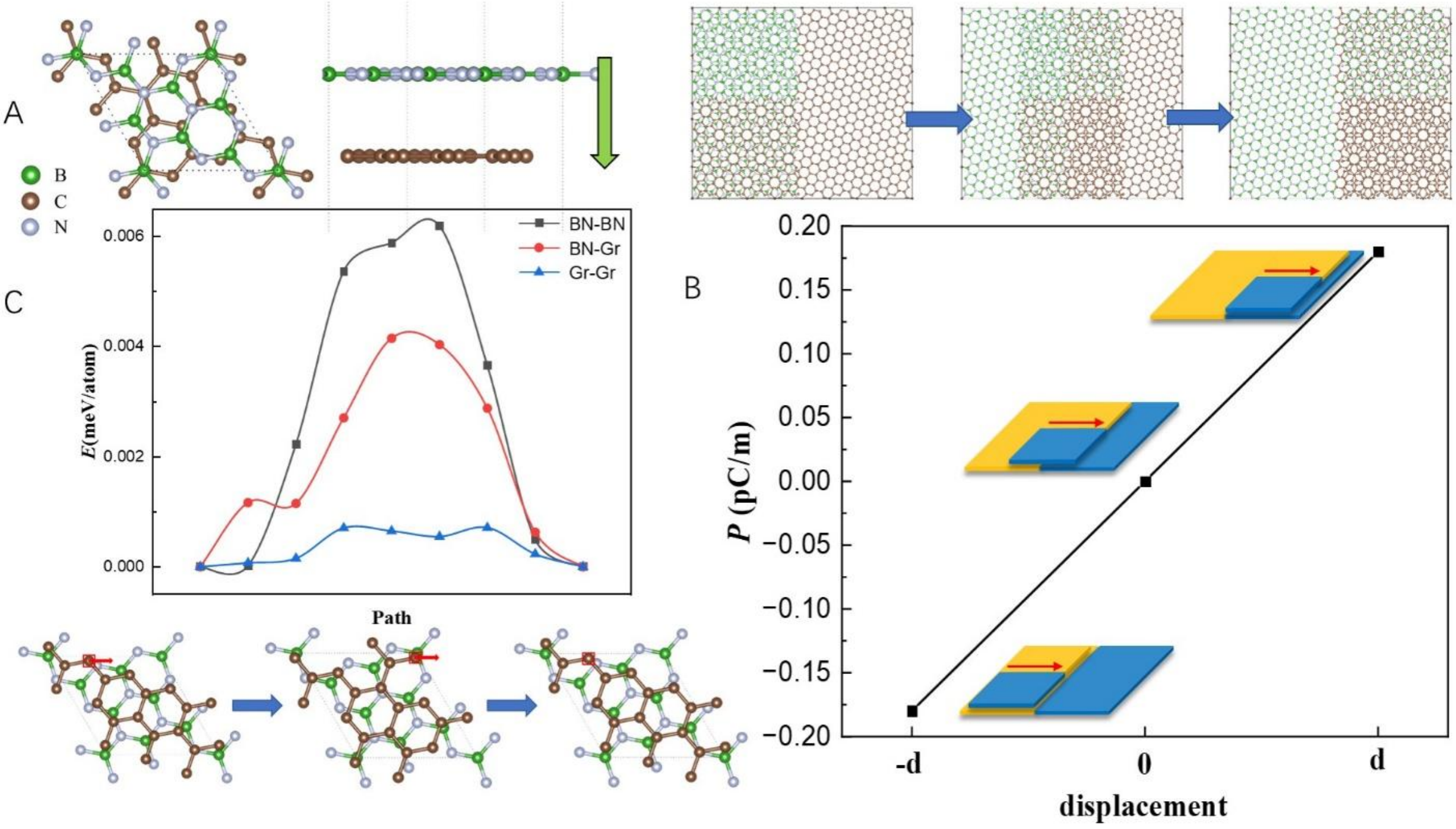

Figure 2. (A) Gr/BN heterobilayer with a large twist angle and a vertical polarization marked by the green arrow. (B) The evolution of vertical polarization for the device based on Gr/BN junction bilayer upon interlayer sliding. (C) Superlubric sliding pathway for BN/Gr, BN/BN and Gr/Gr bilayers.

Table 1 The vertical polarizations of various A/B heterobilayers.

| | Gr/BN | Gr/$MoS_2$ | Gr/$WSe_2$ | Gr/$SnSe_2$ | $MoS_2$/$WSe_2$ | $MoSe_2$/$WS_2$ | $MoS_2$/$MoSe_2$ | $WS_2$/$WSe_2$ |
|---|---|---|---|---|---|---|---|---|
| P(pC/m) | 0.38 | 4.1 | 2.1 | 7.0 | 1.1 | 0.40 | 0.89 | 0.58 |

As mentioned above, the synthesis of many such AB junctions based on transition metal dichalcogenides (TMDs) like Gr/$MoS_2$,[28-30] Gr/$WSe_2$,[31] $MoS_2$/$WSe_2$,[32] $MoSe_2$/$WS_2$, $MoS_2$/$WS_2$, $WS_2$/$WSe_2$, etc.[33], have also been reported, where better candidates for higher polarizations can be filtered. As listed in Table 1, the vertical polarizations of various heterobilayers can be much higher, e.g., the polarizations of Gr/$MoS_2$ and Gr/$SnSe_2$ are more than an order of magnitude higher compared with Gr/BN, also much higher compared with all previous reported sliding ferroelectricity. The Hirshfeld charge analysis of Gr/$SnSe_2$ in Fig. S2 reveals that both interlayer charge transfer and the difference in two Se layers contribute to its large polarization, which also gives rise to a high charge carrier density of $1.63\times10^{13}|e|/cm^2$ in graphene layer. Additionally, the areal interlayer van der Waals binding energies for the bilayers have been all computed in previous study,[49] where the estimated small difference between AA and BB regions favor sliding across the boundary. The superlubricity in twisted graphene/TMD heterobilayers[50, 51] and TMD homobilayers[52] have also been demonstrated in previous reports. The switching voltage for most sliding ferroelectrics are within several volts[5, 9], and given the higher polarizations and barriers reduced by 2-3 orders of magnitudes in our design, the required driving voltage should be greatly reduced.

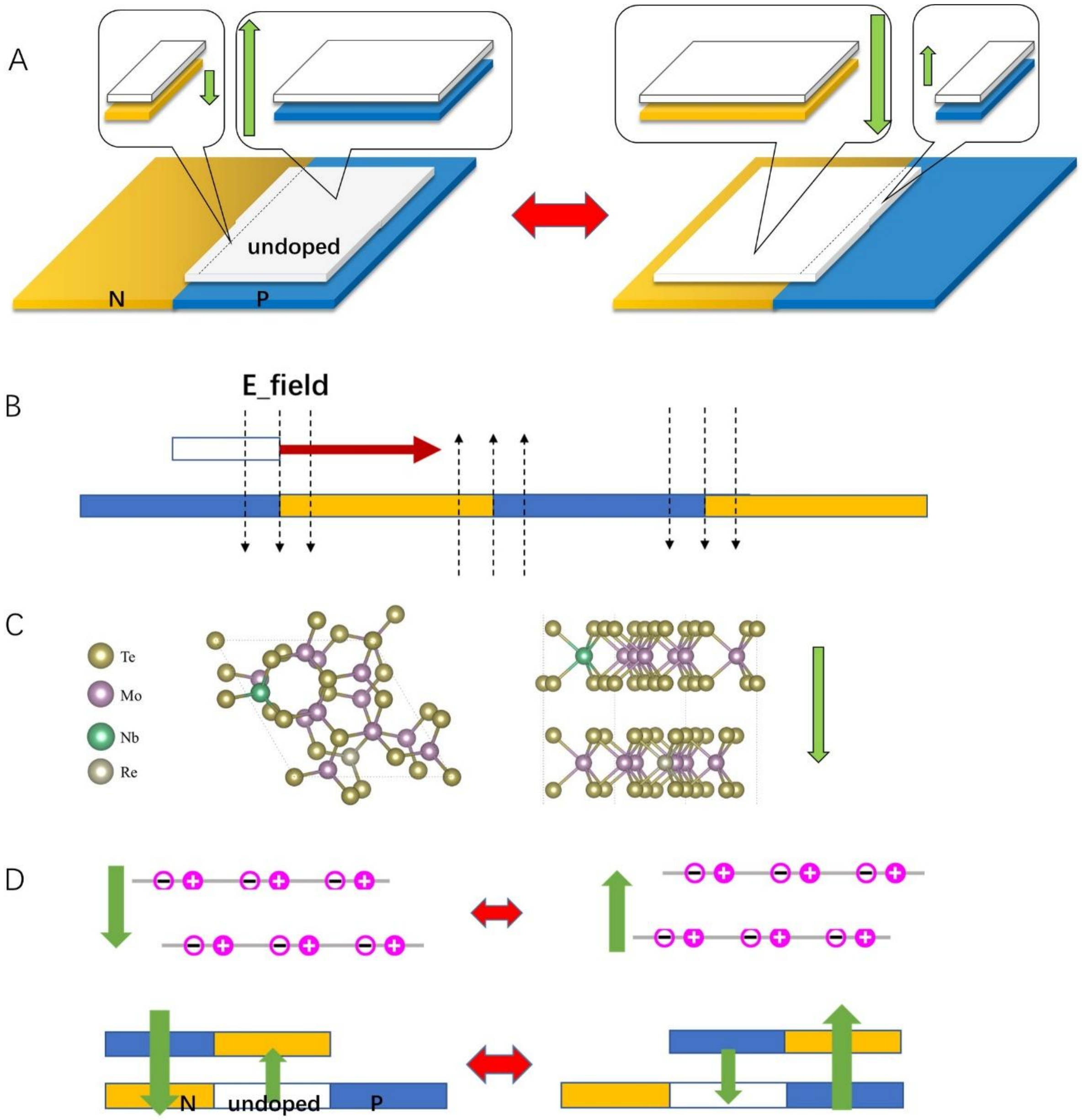

Figure 3. (A) The design of superlubric sliding ferroelectric device based on PN junction monolayer with an undoped flake on the top. (B) The design of long distance superlubric sliding device based on PN superlattice monolayer with an undoped flake on the top, where the acceleration of the flake can be driven by local vertical electric field at the junctions. (C)The Nb-doped/Re-doped $MoTe_2$ twisted bilayer with a vertical polarization marked by green arrow. (D) Comparison of sliding ferroelectric model and superlubric bilayer of PN junction over P-undoped-N junction, where the polarization of P/N region is much larger compared with the P/undoped or N/undoped region in the middle.

Similar principle of design can even be applied to 2D PN junctions based on the same material, since the charge transfers of PN bilayers also give rise to vertical polarizations. Similar bilayer junction structure in Fig. 1B can be adopted, and actually, the top PN junction layer can be replaced by a pristine undoped layer so the fabrication can be greatly facilitated, where vertical polarizations can also be induced by interlayer charge transfer. As shown in Fig. 3A, when a pristine flake is placed on a PN junction flake, the polarization can also be tuned via interlayer sliding, and such superlubric sliding can be driven by electric field as it will be more favorable in energy for the polarization to be aligned along the same direction. To realize superlubric sliding for long distance, the bottom layer can even be a PN superlattice shown in Fig. 3B, where the acceleration of the flake can be driven by local electric field pulse at the junctions. For example, based on previous report[53] of controllable P/N doping of wafer-scale $MoTe_2$ thin films via substitution Mo atoms by Re/Nb atoms, we construct the doped bilayer with a large twist angle of 38.2° in Fig. 3C. When two Mo atoms of the two layers are respectively substituted by Re and Nb atom in each supercell, its polarization is estimated to be 8.5 pC/m. This value can be even enhanced to 13.3 pC/m if the host system is $MoS_2$, and even only 1/35 Mo atoms are substituted, the reduced polarization 2.7 pC/m is still higher compared with most sliding ferroelectrics like bilayer BN. Noting that the $MoS_2$ is also known to be n-doped due to the high-density of sulfur vacancies formed in synthesis, and the polarization can maintain when Re-doping is replaced by S vacancy (see Fig. S3). For the device in Fig. 1B based on PN doped $MoTe_2$, its polarization is tunable within the range [-4.25 pC/m, 4.25 pC/m]. If the design in Fig. 3A with an undoped top flake is adopted, since the polarizations of undoped/N-doped and undoped/P-doped $MoTe_2$ are 2.37 and 0.34 pC/m, the average polarization of the device will be tunable within [-2.37 pC/m, 0.34 pC/m] or [-0.34 pC/m, 2.37 pC/m]. We also note that the polarization is greatly reduced when one layer of the P/N stacking region is replaced by undoped layer. For the typical sliding ferroelectric model in Fig. 3D, if the positive and negative charged ions are respectively replaced by P/N doped nanostrip domains, a design of “super sliding ferroelectric” heterojunction bilayer can be obtained, with uncompensated vertical polarization tunable via long-distance superlubric sliding.

## Discussion

In summary, we propose a design of heterojunctions with continuously tunable vertical polarizations via electrically driven superlubric sliding, rendering series of multi-states for artificial synaptic devices. Compared with typical homobilayer sliding ferroelectrics, the switching barriers of superlubric heterojunctions are greatly reduced and the switching mechanism is even much more unconventional, where polarizations are generated by charge transfer of heterolayers, and controlled by tuning the ratio of different stacking regions via sliding of much longer distance. Moreover, such superlubric sliding electrically driven by low vertical voltage is much more convenient compared with sliding mechanically driven by tips in previous studies. Since the design can be realized based on various systems even including PN junctions, out work should stimulate experimental efforts leading to breakthroughs in both fields of ferroelectricity and superlubricity.

## Methods

The heterojunction systems in our study cannot be simulated in the periodic model, so they are divided into different stacking regions, and their properties are calculated by density functional theory (DFT) methods implemented in the Vienna Ab initio Simulation Package (VASP 5.4) code. [54, 55] The exchange-correlation potential was treated in the Perdew-Burke Ernzerhof (PBE)[56] form under generalized gradient approximation (GGA). [57] A van der Waals correction DFT-D3 with Becke-Johnson damping function has been adopted, [58] which could provide a good description of van der Waals interactions. For structural optimization, the convergence criteria for energy was set to be $10^{-6}$ eV and the plane wave cutoff was set to be 400 eV. The convergence criteria for energy and force are set to be $10^{-6}$ eV and 0.01eV/Å for structure optimization. The Brillouin zone was sampled by 5 × 5 × 1 k points using the Monkhorst–Pack scheme[59]. The vertical polarizations and sliding pathways were calculated using dipole correction method[60] and nudged elastic band (NEB) method, [61] where the sliding barriers of incommensurate interface approximately treated in a large unitcell may be overestimated.

## Resource availability

### Lead contact

Requests for further information and resources should be directed to and will be fulfilled by the lead contact, Menghao Wu (wmh1987@hust.edu.cn).

Materials availability

The work did not generate new materials.

Data and code availability

Any additional information required to reanalyze the data reported in this paper is available from the lead contact upon request.

Acknowledgements

This work is supported by National Natural Science Foundation of China (Nos. 12574263). We thank Prof. Deli Peng and Prof. Wei Cao for helpful discussions.

Author contributions

Conceptualization, M. W.; methodology, P. S.; investigation, P.S., and M. W.; writing – original draft, P.S., and M. W.; writing – review & editing, M. W.; supervision, M. W.

Declaration of interest

The authors declare no competing interests.

Supplemental Information

References

1 Abrahams SC. Systematic prediction of new ferroelectrics on the basis of structure. *Ferroelectrics* 1990; **104**:37-50.
2 Li L, Wu M. Binary Compound Bilayer and Multilayer with Vertical Polarizations: Two-Dimensional Ferroelectrics, Multiferroics, and Nanogenerators. *ACS Nano* 2017; **11**:6382-6388.
3 Wu M, Li J. Sliding ferroelectricity in 2D van der Waals materials: Related physics and future opportunities. *Proceedings of the National Academy of Sciences* 2021; **118**:e2115703118.
4 Vizner Stern M, Waschitz Y, Cao W *et al.* Interfacial ferroelectricity by van der Waals sliding. *Science* 2021; **372**:1462.
5 Yasuda K, Wang X, Watanabe K, Taniguchi T, Jarillo-Herrero P. Stacking-engineered ferroelectricity in bilayer boron nitride. *Science* 2021; **372**:1458.
6 Fei Z, Zhao W, Palomaki TA *et al.* Ferroelectric switching of a two-dimensional metal. *Nature*

2018; **560**:336-339.
7 Yeo Y, Sharaby Y, Roy N *et al.* Polytype switching by super-lubricant van der Waals cavity arrays. *Nature* 2025; **638**:389-393.
8 Rogée L, Wang L, Zhang Y *et al.* Ferroelectricity in untwisted heterobilayers of transition metal dichalcogenides. *Science* 2022; **376**:973-978.
9 Wang X, Yasuda K, Zhang Y *et al.* Interfacial ferroelectricity in rhombohedral-stacked bilayer transition metal dichalcogenides. *Nature Nanotechnology* 2022; **17**:367-371.
10 Weston A, Castanon EG, Enaldiev V *et al.* Interfacial ferroelectricity in marginally twisted 2D semiconductors. *Nature Nanotechnology* 2022; **17**:390-395.
11 Wan Y, Hu T, Mao X *et al.* Room-Temperature Ferroelectricity in $1{\mathrm{T}}^{\ensuremath{'}}$-${\mathrm{ReS}}_{2}$ Multilayers. *Physical Review Letters* 2022; **128**:067601.
12 Deb S, Cao W, Raab N *et al.* Cumulative polarization in conductive interfacial ferroelectrics. *Nature* 2022; **612**:465-469.
13 Jindal A, Saha A, Li Z *et al.* Coupled ferroelectricity and superconductivity in bilayer Td-MoTe2. *Nature* 2023; **613**:48-52.
14 Tsymbal EY. Two-dimensional ferroelectricity by design. *Science* 2021; **372**:1389-1390.
15 Yasuda K, Zalys-Geller E, Wang X *et al.* Ultrafast high-endurance memory based on sliding ferroelectrics. *Science* 2024; **385**:53-56.
16 Bian R, He R, Pan E *et al.* Developing fatigue-resistant ferroelectrics using interlayer sliding switching. *Science* 2024; **385**:57-62.
17 Fan A, Zhang Q, Yang Z *et al.* Tailored sliding ferroelectricity for ultrahigh fatigue resistance in stacked trilayer MoS2 crystals. *Science Advances*; **11**:eadx8192.
18 Bai Y, Yu Z, Guan Z *et al.* Sub-nanosecond polarization switching with anomalous kinetics in vdW ferroelectric WTe2. *Nature Communications* 2025; **16**:7221.
19 Liu Z, Yang J, Grey F *et al.* Observation of Microscale Superlubricity in Graphite. *Physical Review Letters* 2012; **108**:205503.
20 Vazirisereshk MR, Ye H, Ye Z *et al.* Origin of Nanoscale Friction Contrast between Supported Graphene, MoS2, and a Graphene/MoS2 Heterostructure. *Nano Letters* 2019; **19**:5496-5505.
21 Song Y, Mandelli D, Hod O, Urbakh M, Ma M, Zheng Q. Robust microscale superlubricity in graphite/hexagonal boron nitride layered heterojunctions. *Nature Materials* 2018; **17**:894-899.
22 Liao M, Nicolini P, Du L *et al.* Ultra-low friction and edge-pinning effect in large-lattice-mismatch van der Waals heterostructures. *Nature Materials* 2022; **21**:47-53.
23 Yang Z, Wu M. Superlubric sliding ferroelectricity. *Applied Physics Reviews* 2025; **12**:021417.
24 Huang X, Xiang X, Li C *et al.* Electrostatic in-plane structural superlubric actuator. *Nature Communications* 2025; **16**:493.
25 Liu Z, Ma L, Shi G *et al.* In-plane heterostructures of graphene and hexagonal boron nitride with controlled domain sizes. *Nature Nanotechnology* 2013; **8**:119-124.
26 Sutter P, Huang Y, Sutter E. Nanoscale Integration of Two-Dimensional Materials by Lateral Heteroepitaxy. *Nano Letters* 2014; **14**:4846-4851.
27 Park J, Lee J, Liu L *et al.* Spatially resolved one-dimensional boundary states in graphene–hexagonal boron nitride planar heterostructures. *Nature Communications* 2014; **5**:5403.
28 Zhao M, Ye Y, Han Y *et al.* Large-scale chemical assembly of atomically thin transistors and circuits. *Nature Nanotechnology* 2016; **11**:954-959.

29 Guimarães MHD, Gao H, Han Y *et al.* Atomically Thin Ohmic Edge Contacts Between Two-Dimensional Materials. *ACS Nano* 2016; **10**:6392-6399.
30 Subramanian S, Campbell QT, Moser SK *et al.* Photophysics and Electronic Structure of Lateral Graphene/MoS2 and Metal/MoS2 Junctions. *ACS Nano* 2020; **14**:16663-16671.
31 Tang H-L, Chiu M-H, Tseng C-C *et al.* Multilayer Graphene–WSe2 Heterostructures for WSe2 Transistors. *ACS Nano* 2017; **11**:12817-12823.
32 Li M-Y, Shi Y, Cheng C-C *et al.* Epitaxial growth of a monolayer WSe2-MoS2 lateral p-n junction with an atomically sharp interface. *Science* 2015; **349**:524-528.
33 Zhang Z, Chen P, Duan X, Zang K, Luo J, Duan X. Robust epitaxial growth of two-dimensional heterostructures, multiheterostructures, and superlattices. *Science* 2017; **357**:788-792.
34 Ren Y, Wu M. 0D/1D organic ferroelectrics/multiferroics for ultrahigh density integration: Helical hydrogen-bonded chains, multi-mode switching, and proton synaptic transistors. *JChemPhys* 2021; **154**:044705.
35 Tian B, Liu L, Yan M *et al.* A Robust Artificial Synapse Based on Organic Ferroelectric Polymer. *Advanced Electronic Materials* 2019; **5**:1800600.
36 Juma IG, Kim G, Jariwala D, Behura SK. Direct growth of hexagonal boron nitride on non-metallic substrates and its heterostructures with graphene. *iScience* 2021; **24**.
37 Ru G, Qi W, Tang K, Wei Y, Xue T. Interlayer friction and superlubricity in bilayer graphene and MoS2/MoSe2 van der Waals heterostructures. *Tribology International* 2020; **151**:106483.
38 Mandelli D, Leven I, Hod O, Urbakh M. Sliding friction of graphene/hexagonal –boron nitride heterojunctions: a route to robust superlubricity. *Scientific Reports* 2017; **7**:10851.
39 Lebedev AV, Lebedeva IV, Popov AM, Knizhnik AA. Stacking in incommensurate graphene/hexagonal-boron-nitride heterostructures based on ab initio study of interlayer interaction. *Physical Review B* 2017; **96**:085432.
40 Yang Q, Wu M, Li J. Origin of Two-Dimensional Vertical Ferroelectricity in WTe2 Bilayer and Multilayer. *The Journal of Physical Chemistry Letters* 2018; **9**:7160-7164.
41 Sharma P, Xiang F-X, Shao D-F *et al.* A room-temperature ferroelectric semimetal. *Science Advances* 2019; **5**:eaax5080.
42 Santos EJG, Kaxiras E. Electric-Field Dependence of the Effective Dielectric Constant in Graphene. *Nano Letters* 2013; **13**:898-902.
43 Li LH, Tian T, Cai Q, Shih C-J, Santos EJG. Asymmetric electric field screening in van der Waals heterostructures. *Nature Communications* 2018; **9**:1271.
44 Lin F, Xuan X, Cao Z *et al.* Room temperature ferroelectricity in monolayer graphene sandwiched between hexagonal boron nitride. *Nature Communications* 2025; **16**:1189.
45 Roy N, Ying P, Atri SS *et al.* Switching graphitic polytypes in elastically coupled cavities. *Nature Nanotechnology* 2026.
46 Zheng Z, Ma Q, Bi Z *et al.* Unconventional ferroelectricity in moiré heterostructures. *Nature* 2020; **588**:71-76.
47 Ravindran P, Vidya R, Kjekshus A, Fjellvåg H, Eriksson O. Theoretical investigation of magnetoelectric behavior in $\mathrm{Bi}\mathrm{Fe}{\mathrm{O}}_{3}$. *Physical Review B* 2006; **74**:224412.
48 Shelke V, Mazumdar D, Srinivasan G *et al.* Reduced Coercive Field in BiFeO3 Thin Films Through Domain Engineering. *Advanced Materials* 2011; **23**:669-672.
49 Tang K, Qi W, Wei Y, Ru G, Liu W. High-Throughput Calculation of Interlayer van der Waals

Forces Validated with Experimental Measurements. *Research*; **2022**.
50 Wang L, Zhou X, Ma T *et al.* Superlubricity of a graphene/MoS2 heterostructure: a combined experimental and DFT study. *Nanoscale* 2017; **9**:10846-10853.
51 Büch H, Rossi A, Forti S, Convertino D, Tozzini V, Coletti C. Superlubricity of epitaxial monolayer WS2 on graphene. *Nano Research* 2018; **11**:5946-5956.
52 Onodera T, Morita Y, Nagumo R *et al.* A Computational Chemistry Study on Friction of h-MoS2. Part II. Friction Anisotropy. *The Journal of Physical Chemistry B* 2010; **114**:15832-15838.
53 Pan Y, Jian T, Gu P *et al.* Precise p-type and n-type doping of two-dimensional semiconductors for monolithic integrated circuits. *Nature Communications* 2024; **15**:9631.
54 Kresse G, Furthmüller J. Efficient iterative schemes for ab initio total-energy calculations using a plane-wave basis set. *Physical Review B* 1996; **54**:11169-11186.
55 Kresse G, Furthmüller J. Efficiency of ab-initio total energy calculations for metals and semiconductors using a plane-wave basis set. *Computational Materials Science* 1996; **6**:15-50.
56 Perdew JP, Burke K, Ernzerhof M. Generalized Gradient Approximation Made Simple. *Physical Review Letters* 1996; **77**:3865-3868.
57 Perdew JP, Chevary JA, Vosko SH *et al.* Atoms, molecules, solids, and surfaces: Applications of the generalized gradient approximation for exchange and correlation. *Physical Review B* 1992; **46**:6671-6687.
58 Grimme S, Ehrlich S, Goerigk L. Effect of the damping function in dispersion corrected density functional theory. *Journal of Computational Chemistry* 2011; **32**:1456-1465.
59 Monkhorst HJ, Pack JD. Special points for Brillouin-zone integrations. *Physical Review B* 1976; **13**:5188-5192.
60 Neugebauer J, Scheffler M. Adsorbate-substrate and adsorbate-adsorbate interactions of Na and K adlayers on Al(111). *Physical Review B* 1992; **46**:16067-16080.
61 Henkelman G, Uberuaga BP, Jónsson H. A climbing image nudged elastic band method for finding saddle points and minimum energy paths. *The Journal of Chemical Physics* 2000; **113**:9901-9904.